\documentclass[pra,twocolumn,showpacs]{revtex4}
\makeatletter
\newcommand{\@ptsize}{0}
\makeatother
\usepackage{natbib}
\usepackage{epsfig,emp,latexsym,amsmath}

\newcommand{\ket}[1]{\left|#1\right>}

\newcommand{\nix}[1]{}
\newcommand{\C}{{\mathbf{C}}}

\newcommand{\diag}{\mbox{diag}}

\newcommand{\onemat}{{\mathbf{1}}}
\newcommand{\qed}{\mbox{$\Box$}}
\newcommand{\proof}{\textit{Proof.}\ }

\newtheorem{lemma}{Lemma}
\newtheorem{theorem}[lemma]{Theorem}

\newcommand{\mysection}{\section}

\begin{document}

\title{Engineering Functional Quantum Algorithms}
\author{Andreas Klappenecker}
\email{klappi@cs.tamu.edu}
\affiliation{Department of Computer Science, Texas A\&M University, 
College Station, TX 77843-3112, USA}
\thanks{Research partly supported by NSF grant EIA 0218582, and a Texas A\&M TITF grant.} 

\author{Martin R\"otteler}
\email{roettele@ira.uka.de}
\affiliation{Universit\"at Karlsruhe, 
Insitut f\"ur Algorithmen und Kognitive Systeme, 
Forschungsgruppe Professor Beth, 
D-76128 Karlsruhe, Germany} 
\thanks{Research supported by EC grant IST-1999-10596 (Q-ACTA).} 

\begin{abstract}
\noindent 
Suppose that a quantum circuit with $K$ elementary gates is known for
a unitary matrix~$U$, and assume that $U^m$ is a scalar matrix for
some positive integer $m$. We show that a function of $U$ can be
realized on a quantum computer with at most $O(mK+m^2\log m)$
elementary gates.  The functions of $U$ are realized by a generic
quantum circuit, which has a particularly simple structure. Among
other results, we obtain efficient circuits for the fractional Fourier
transform.
\end{abstract}
\maketitle
\empprelude{input qcg; prologues := 0;}
\begin{empfile}


\noindent 
Let $U$ be a unitary matrix, $U\in {\mathcal U}(2^n)$. Suppose that a
fast quantum algorithm is known for $U$, which is given
by a factorization of the form
\begin{equation}\label{eq:factorization}
 U = U_{1}U_2\cdots U_K,
\end{equation}
where the unitary matrices $U_i$ are realized by controlled-not gates
or by single qubit gates~\cite{BBCshort:95}.  We are interested in the
following question:
\begin{quote}
\textsl{Are there efficient quantum algorithms 
for unitary matrices, which are functions of~$U$?}
\end{quote}
The question is puzzling, because the knowledge of the
factorization~(\ref{eq:factorization}) of $U$ does not seem to be of
much help in finding similar factorizations for, say, $V=U^{1/3}$.
The purpose of this letter is to give an answer to the above question
for a wide range of unitary matrices $U$.

Our solution to this problem is based on a generic circuit which
implements arbitrary functions of $U$, assuming that $U^m$ is a scalar
matrix for some positive integer $m$.  If $m$ is small, then our
method provides an efficient quantum circuit for~$V$. \nix{An appealing
aspect of our approach is that it provides a uniform solution to the
design of a wide variety of quantum algorithms.}

\textit{Notations.}  We denote by $\mathcal{U}(m)$ the group of
unitary $m\times m$ matrices, by $\onemat$ the identity matrix, and by
$\C$ the field of complex numbers.

\mysection{Preliminaries}
\noindent We recall some standard material on matrix functions,
see~\cite{ferrar51,gantmacher60,horn91} for more details.  Let $U$ be
a unitary matrix. The spectral theorem states that $U$ is unitarily
equivalent to a diagonal matrix $D$, that is, $U=T DT^\dagger$ for
some unitary matrix $T$.  The elements $\lambda_i$ on the diagonal of
$D=\mbox{diag}(\lambda_1, \dots, \lambda_{2^n})$ are the eigenvalues
of $U$.

Let $f$ be any function of complex scalars such that its domain
contains the eigenvalues $\lambda_i$, $1\le i\le 2^n$.   
The matrix function
$f(U)$ is then defined by
$$ f(U) = T\mbox{diag}(f(\lambda_1),\dots, f(\lambda_{2^n}))\,
T^\dagger,$$
where $T$ denotes the diagonalizing matrix of~$U$, as above.

Notice that any two scalar functions $f$ and $g$, which take the same
values on the spectrum of $U$, yield the same matrix value
$f(U)=g(U)$.  In particular, one can find an interpolation polynomial
$g$, which takes the same values as $f$ on the eigenvalues
$\lambda_i$.  It is possible to assume that the degree of $g$ is
smaller than the degree of the minimal polynomial of $U$.
In other words, $V=f(U)$ can be expressed by a linear combination of
integral powers of the matrix $U$,
\begin{equation}\label{eq:lk}
V= f(U) = \sum_{i=0}^{m-1} \alpha_i U^i,
\end{equation}
where $m$ is the degree of the minimal polynomial of the matrix $U$, and $\alpha_i\in \C$ for $i=0,\dots, m-1$. 
In order for $V$ to be unitary, it is necessary and sufficient that
the function $f$ maps the eigenvalues $\lambda_i$ of $U$ to elements
on the unit circle.

\textit{Remark.}  There exist several different definitions for
matrix functions. The relationshop between these definitions is
discussed in detail in~\cite{rinehart55}. We have chosen the most
general definition that allows to express the function values by
polynomials.

\mysection{The Generic Circuit} 
\noindent Let $U$ be a unitary $2^n\times 2^n$
matrix with minimal polynomial of degree $m$. We assume that an
efficient quantum circuit is known for $U$. 
How can we go about implementing the linear combination~(\ref{eq:lk})?
We will use an ancillary system of $\mu$ quantum bits, where $\mu$ is
chosen such that $2^{\mu-1}< m\le 2^{\mu}$ holds. This will allow us
to create the linear combination by manipulating somewhat larger
matrices, which on input $\ket{0}\otimes
\ket{\psi}\in\C^{2^\mu}\otimes \C^{2^n}$ produce the state
$\ket{0}\otimes V\ket{\psi}$. 

We first bring the ancillary system into a superposition of the first
$m$ computational base states, such that an input state $\ket{0}\otimes
\ket{\psi}\in \C^{2^\mu}\otimes \C^{2^n}$ is mapped to the state
\begin{equation}\label{eq:spos} 
\frac{1}{\sqrt{m}}\sum_{i=0}^{m-1} \ket{i}\otimes \ket{\psi}.
\end{equation}
 This
can be done by acting with a $2^\mu\times 2^\mu$ unitary matrix~$B$ on
the ancillary system, where the first column of $B$ is of the form
$1/\sqrt{m}(1,\dots,1,0,\dots,0)^t$.
Efficient implementations of $B$ exist.

Notice that there exists an efficient implementation of the block
diagonal matrix $A=\mbox{diag}(1,U,U^2,\dots,U^{2^{\mu}-1})$. Indeed,
$A$ can be composed of the matrices $U^{2^\eta}$, $0\le \eta<\mu$,
conditioned on the $\mu$ ancillae bits.  The resulting implementation
is shown in Fig.~\ref{fig:diag}.
The state (\ref{eq:spos}) 
is transformed by this circuit into the state
\begin{equation}\label{eq:di} 
\frac{1}{\sqrt{m}}\sum_{i=0}^{m-1} \ket{i}\otimes U^i\ket{\psi}.
\end{equation}

\begin{figure}
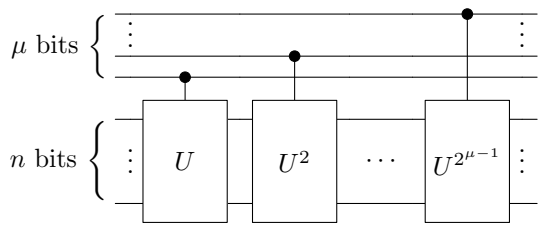

\begin{center}
\begin{emp}(50,50)
 setunit 1.4mm;
 qubits(6); 
 dropwire(1); 

 QCycoord[3] := QCycoord[3] - 1/2QCheight;
 QCycoord[4] := QCycoord[4] - 1/2QCheight;
 label.lft(btex $\mu \mbox{ bits }\bigg\{ $ etex,(QCxcoord,QCycoord[3]+1.3mm));
 label.lft(btex $n \mbox{ bits }\Bigg\{ $ etex,(QCxcoord,QCycoord[1]-5.5mm));

 wires(2mm);
 label(btex $\vdots$ etex, (QCxcoord , QCycoord[0]+1.2QCheight)); 
 label(btex $\vdots$ etex, (QCxcoord, QCycoord[2]+1.2QCheight));
 circuit(2QCheight)(icnd 2, gpos 0,1,btex $U$ etex);
 circuit(2QCheight)(icnd 3, gpos 0,1,btex $U^2$ etex);
 label(btex $\cdots$ etex, (QCxcoord + 1/2QCstepsize, QCycoord[0]+QCheight));
 wires(3/2QCheight);
 circuit(2QCheight)(icnd 4, gpos 0,1,btex $U^{2^{\mu-1}}$ etex);
 label(btex $\vdots$ etex, (QCxcoord , QCycoord[0]+1.2QCheight)); 
 label(btex $\vdots$ etex, (QCxcoord, QCycoord[2]+1.2QCheight));
 wires(2mm);
\end{emp}
\end{center}
\caption{A quantum circuit realizing the block diagonal matrix 
$A=\mbox{diag}(1,U,U^2,\dots,U^{2^{\mu}-1})$. 
}\label{fig:diag}
\end{figure}

In the next step, we let a $2^\mu\times 2^\mu$ matrix $M$ act on the
ancillae bits. We choose $M$ such that the state~(\ref{eq:di}) is mapped to 
\begin{equation}\label{eq:M} 
\frac{1}{\sqrt{m}}\sum_{k=0}^{m-1} \ket{k}\otimes U^kV\ket{\psi}
\end{equation}
It turns out that~$M$ can be realized by a unitary matrix, assuming
that the minimal polynomial of $U$ is of the form $x^m-\tau$, 
$\tau\in \C$. This will be explained in some detail in the next section.

We apply the inverse $A^\dagger$ of the block diagonal
matrix~$A$. This transforms the state~(\ref{eq:M}) to
\begin{equation}\label{eq:Ainv}
\frac{1}{\sqrt{m}}\sum_{k=0}^{m-1} \ket{k}\otimes V\ket{\psi}.
\end{equation}
We can clean up the ancillae bits by applying the $2^\mu\times 2^\mu$
matrix~$B^\dagger$. This yields then the output state 
\begin{equation}\label{eq:fin}
\ket{0} \otimes V\ket{\psi}= \ket{0}\otimes f(U)\ket{\psi}.
\end{equation}
The steps from the input state $\ket{0}\otimes \ket{\psi}$ to the
final output state $\ket{0}\otimes V\ket{\psi}$ are illustrated in
Fig.~\ref{fig:final} for the case $\mu=2$. 
\begin{figure}[ht]
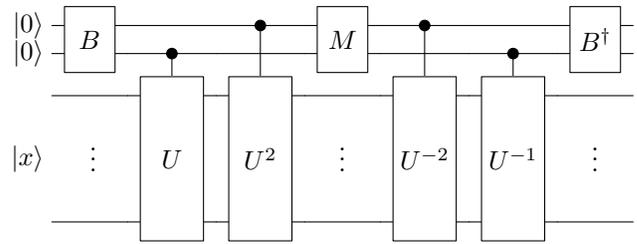

\begin{center}
\begin{emp}(50,50)
  setunit 1.4mm; qubits(6); QCycoord[5] := QCycoord[5]-1/3QCheight;
  ysave := QCycoord[2]-1/2QCheight; label.lft(btex $|0\rangle$
  etex,(QCxcoord,QCycoord[5])); label.lft(btex $|0\rangle$
  etex,(QCxcoord,QCycoord[4])); label.lft(btex $|x\rangle$
  etex,(QCxcoord,ysave)); dropwire(1,2); ysave := ysave + 1mm;
  label(btex $\vdots$ etex,(QCxcoord+5.5mm,ysave));

  circuit(1.2*QCheight)(gpos 2,3,btex $B$ etex);
  circuit(1.5*QCheight)(icnd 2,gpos 0,1,btex $U$ etex);
  circuit(1.5*QCheight)(icnd 3,gpos 0,1,btex $U^2$ etex);

  label(btex $\vdots$ etex,(QCxcoord+5.0mm,ysave));
  circuit(1.2*QCheight)(gpos 2,3,btex $M$ etex);

  circuit(1.5*QCheight)(icnd 3,gpos 0,1,btex $U^{-2}$ etex);
  circuit(1.5*QCheight)(icnd 2,gpos 0,1,btex $U^{-1}$ etex);
  label(btex $\vdots$ etex,(QCxcoord+5.0mm,ysave));
  circuit(1.2*QCheight)(gpos 2,3,btex $B^\dagger$ etex);
\end{emp}
\end{center}
\caption{Generic circuit realizing a linear combination $V$. The case $\mu=2$ is shown.}\label{fig:final}
\end{figure}

The following theorem gives an upper bound on the complexity of the
method. We use the number of elementary gates (that is, the number of
single qubit gates and controlled-not gates) as a measure of
complexity.

\begin{theorem}\label{theorem} 
Let $U$ be a $2^n\times 2^n$ unitary matrix with 
minimal polynomial $x^m-\tau$, $\tau\in \C$. 
Suppose that there exists a quantum algorithm for $U$ using $K$
elementary gates. Then a unitary matrix $V=f(U)$ can be realized with at
most $O(mK+m^2\log m)$ elementary operations. 
\end{theorem}
\proof A matrix acting on $\mu\in O(\log m)$ qubits can be realized
with at most $O(m^2\log m)$ elementary
operations, cf.~\cite{BBCshort:95}. Therefore, the matrices $B, B^\dagger$, and
$M$ can be realized with a total of at most $O(3m^2\log m)$
operations.

If $K$ operations are needed to implement $U$, then at most $14K$
operations are needed to implement $\Lambda_1(U)$, the operation $U$
controlled by a single qubit.  The reason is that a doubly controlled
NOT gate can be implemented with 14 elementary
gates~\cite{divincenzo98}, and a controlled single qubit gate can be
implemented with six or fewer elementary gates~\cite{BBCshort:95}.

We observe that $2^\mu-1$ copies of $\Lambda_1(U)$ suffice to
implement $A$.  Indeed, we certainly can implement
$\Lambda_1(U^{2^k})$ by a sequence of $2^k$ circuits $\Lambda_1(U)$. 
This bold implementation yields the estimate for $A$. Typically, we will be able to find much more efficient implementations. Anyway, we can conclude that 
$A$ and $A^\dagger$ can both
be implemented by at most $14(2^\mu-1)K\in O(14mK)$ operations.
Combining our counts yields the result.~\qed

\mysection{Unitarity of the matrix $M$}\label{sec:magical}
\noindent It remains to show that the state~(\ref{eq:di}) can be
transformed into the state~(\ref{eq:M}) by acting with a unitary
matrix $M$ on the system of $\mu$ ancillae qubits. This is the crucial
step in the previously described method. 

Let $U$ be a unitary matrix with a minimal polynomial of degree $m$. 
A unitary matrix $V=f(U)$ can then be 
represented by a linear combination
\begin{equation}\label{eq:insight}
V=\sum_{i=0}^{m-1} \alpha_i U^i.
\end{equation}
We will motivate the construction of the matrix $M$ by examining in
some detail the resulting linear combinations of the matrices $U^kV$.
From (\ref{eq:insight}), we obtain 
\begin{equation}\label{eq:insight2} 
U^kV=\sum_{i=0}^{m-1} \alpha_i U^{i+k}.
\end{equation}
Suppose that the minimal polynomial of $U$ is of the form 
$m(x)=x^m-g(x)$, with $g(x)=\sum_{i=0}^{m-1} g_ix^i$. 
The right hand side of (\ref{eq:insight2}) can be reduced to a polynomial in $U$ of degree less than $m$ using the relation 
$U^m=g(U)$: 
$$ U^kV=\sum_{i=0}^{m-1}\beta_{ki} U^i.$$
The coefficients $\beta_{ki}$ are explicitly given by 
$$ (\beta_{k0},\beta_{k1},\dots,\beta_{k(m-1)})=
(\alpha_0,\alpha_1,\dots,\alpha_{m-1})P^k
$$ 
where $P$ denotes the companion matrix of $m(x)$, that is,
$$
P = \left(\begin{array}{ccccc}
0 & 1 & 0 &\cdots & 0\\
0 & 0 & 1 & \cdots & 0\\
\vdots & \vdots  & \vdots & \ddots& \vdots\\
0 & 0 & 0 & \cdots & 1 \\
g_0 & g_1 & g_2 & \dots & g_{m-1}
	  \end{array}
\right).
$$
The $2^\mu\times 2^\mu$ matrix $M$ is defined by 
$$ M = \left(\begin{array}{cc}
C & 0\\
0 & \onemat
\end{array}
\right),
$$
where $C=(\beta_{ki})_{k,i=0,\dots,m-1}$, and $\onemat$ is a $(2^\mu-m)\times (2^\mu-m)$ identity matrix.  Under the assumptions of
Theorem~\ref{theorem}, it turns out that the matrix $M$ is 
unitary. Before proving this claim,
let us formally check that the matrix $M$ transforms the state~(\ref{eq:di})
into the state (\ref{eq:M}). If we apply the matrix $M$ to the
ancillary system, then we obtain from (\ref{eq:di}) the state
\begin{equation*}
\begin{split}
\frac{1}{\sqrt{m}}\sum_{i=0}^{m-1} M\ket{i}& \otimes U^i\ket{\psi} =
\frac{1}{\sqrt{m}}\sum_{k,i=0}^{m-1} \beta_{ki}\ket{k}\otimes U^i\ket{\psi}\\
&=\frac{1}{\sqrt{m}}\sum_{k=0}^{m-1}\ket{k}\otimes
\sum_{i=0}^{m-1}\beta_{ki}U^i\ket{\psi}\\
{} &=\frac{1}{\sqrt{m}}\sum_{k=0}^{m-1}\ket{k}\otimes  U^kV\ket{\psi}
\end{split}
\end{equation*}
which coincides with (\ref{eq:M}), as claimed.

\begin{lemma}
Let $U$ be a unitary matrix with minimal polynomial
$m(x)=x^m-\tau$. Let $V$ be a matrix satisfying (\ref{eq:lk}). 
If $V$ is unitary, then $M$ is unitary. 
\end{lemma}

\proof It suffices to show that the matrix $C$ is unitary. 
Notice that the assumption on the minimal polynomial $m(x)$ implies that 
$C$ is of the form 
$$ C = \left(
\begin{array}{ccccc}
\alpha_0 & \alpha_1 & \cdots & \alpha_{m-2}& \alpha_{m-1}\\
\tau\alpha_{m-1} & \alpha_0 & \cdots & \alpha_{m-3}& \alpha_{m-2}\\
\ddots & \ddots &  & \ddots & \ddots\\
\tau\alpha_1 & \tau\alpha_2 & \cdots & \tau \alpha_{m-1} & \alpha_0
\end{array}
\right),
$$ 
that is, $C$ is obtained from a circulant matrix by multiplying every
entry below the diagonal by $\tau$. In other words, we have 
$$ C = \Big( [\tau]_{i>j}\, \alpha_{j-i\bmod m}\Big)_{i,j=0,\dots,m-1}$$
where $[\tau]_{i>j}=\tau$ if $i>j$, and $[\tau]_{i>j}=1$ otherwise. 

Note that the inner product of row $a$ with row $b$ of matrix
$C$ is the same as the inner product of row $a+1$ with row
$b+1$. Thus, to prove the unitarity of $C$, it suffices to show that 
\begin{equation}\label{eq:inner}
\delta_{a,0}\stackrel{!}{=} \langle \mbox{row a}|\mbox{row 0}\rangle =
\sum_{j=0}^{a-1} \overline{\tau}\,\overline{\alpha_{j-a}}\alpha_j + 
\sum_{j=a}^{m-1} \overline{    \alpha_{j-a}}\alpha_j
\end{equation}
holds, where $\delta_{a,0}$ denotes the Kronecker delta and the
indices of $\alpha$ are understood modulo $m$. 

Consider the equation
\begin{equation}\label{eq:compare}
\onemat = V^\dagger V = 
\left(\sum_{i=0}^{m-1}\overline{\alpha_i} U^{-i}\right)
\left(\sum_{i=0}^{m-1}\alpha_i U^i\right)
\end{equation}
The right hand side can be simplified to a polynomial in $U$ of degree
less than $m$ using the identity $\overline{\tau}\, U^m=\onemat$. The
coefficient of $U^a$ in (\ref{eq:compare}) is exactly the right hand
side of equation~(\ref{eq:inner}).  Since the minimal polynomial of
$U$ is of degree $m$, it follows that the matrices $U^0,
U^1,\dots,U^{m-1}$ are linearly independent.  Thus, comparing
coefficients on both sides of equation~(\ref{eq:compare})
shows~(\ref{eq:inner}). Hence the rows of $C$ are pairwise orthogonal
and of unit norm.~\qed
\medskip

\textit{A Simple Example.} Let $F_n$ be the discrete Fourier
transform matrix 
$$F_n=2^{-n/2}(\exp(-2\pi i\,
{k\ell}/2^n))_{k,\ell=0,\dots,2^n-1},$$ with $i^2=-1$.  Recall that the
Cooley-Tukey decomposition yields a fast quantum algorithm, which
implements $F_n$ with $O(n^2)$ elementary operations. 
The minimal polynomial of $F_n$ is $x^4-1$ if $n\ge 3$. 
Thus, any unitary matrix~$V$, which is a function of $F_n$, 
can be realized with $O(n^2)$ operations. 

For instance, if $n\ge 3$, then the fractional power 
$F_n^x$, $x\in \mathbf{R}$, can be expressed by
$$F_n^x=\alpha_0(x)I+\alpha_1(x) F_n + \alpha_2(x) F_n^2 + \alpha_3(x) F_n^3,$$ 
where the coefficients $\alpha_i(x)$ are given by (cf.~\cite{santhanam96}): 
$$ 
\begin{array}{ll}
\alpha_0(x) = \frac{1}{2}(1+e^{ix})\cos x, &\alpha_1(x) = \frac{1}{2}(1-ie^{ix})\sin x,\\[1ex]
\alpha_2(x) = \frac{1}{2}(-1+e^{ix})\cos x, &\alpha_3(x) = 
\frac{1}{2}(-1-ie^{ix})\sin x.
\end{array}
$$
In this case, $F_n^x$ is realized by the circuit in
Fig.~\ref{fig:final} with $U=F_n$ and
$M=(\alpha_{j-i}(x))_{i,j=0,\dots,3}$.  
The circuit can be implemented with $O(n^2)$ operations.

\mysection{Limitations} 
\noindent The previous sections showed that a unitary matrix $f(U)$
can be realized by a linear combination of the powers
$U^i$, $0\le i<m$, if the minimal polynomial $m(x)$ of $U$ is
of the form $x^m-\tau$, $\tau \in \C$.  One might wonder whether the
restriction to minimal polynomials of this form is really necessary.
The next lemma explains why we had this limitation: 

\begin{lemma} Let  $U$ be a unitary matrix with minimal polynomial 
$m(x)=x^m-g(x)$, $\deg g(x)<m$. If $g(x)$ is not a constant, 
then the matrix $M$ is in general not unitary.
\end{lemma}
\proof Suppose that $g(x)=\sum_{i=0}^{m-1} g_ix^i$. We may choose 
for instance $V=U^m=g(U)$. 
Then the norm of first row in $M$ is
greater than $1$.  Indeed, we 
can calculate this norm to be
$|g_0|^2+|g_1|^2+\cdots+|g_{m-1}|^2$. However, $|g_0|^2=1$, because
$g_0$ is a product of eigenvalues of $U$. By assumption, there is
another nonzero coefficient $g_i$, which proves the result.~\qed

\mysection{Extensions}
\noindent 
We describe in this section one possibility to extend our approach to
a larger class of unitary matrices $U$.  We assumed so far that 
$f(U)$ is realized by a linear combination~(\ref{eq:lk}) of
\textsl{linearly independent} matrices $U^i$. The exponents were
restricted to the range $0\le i< m$, where $m$ is degree of the
minimal polynomial of $U$. We can circumvent the problem indicated in the
previous section by allowing~$m$ to be larger than the degree
of the minimal polynomial.

\begin{theorem} Let $U\in \mathcal{U}(2^n)$ be a unitary matrix
such that $U^m$ is a scalar matrix for some positive integer $m$.
Suppose that there exists a quantum circuit which implements $U$
with $K$ elementary gates. Then a unitary matrix $V=f(U)$ can be realized
with $O(mK+m^2\log m)$ elementary operations.
\end{theorem}
\proof By assumption, $U^m=\tau \onemat$ for some $\tau\in \C$.  This
means that the minimal polynomial $m(x)$ of $U$ divides the polynomial
$x^m-\tau$, that is, $x^m-\tau=m(x)m_2(x)$ for some $m_2(x)\in \C[x]$.  

We may assume without loss of generality that the function $f$ is
defined at all roots of $x^m-\tau$. Indeed, we can replace $f$ by an
interpolation polynomial $g$ satisfying $f(U)=g(U)$ if this is necessary.

Choose any unitary matrix $A\in U(2^n)$ with minimal polynomial
$m_2(x)$.  The minimal polynomial of the block diagonal matrix
$U_A=\diag(U,A)$ is $x^m-\tau$, the least common multiple of the
polynomials $m(x)$ and $m_2(x)$. Express $f(U_A)$ by powers of the
block diagonal matrix $U_A$:
\begin{equation}\label{eq:aux} 
f(U_A)=\diag(f(U),f(A)) =\sum_{i=0}^{m-1} \alpha_i
\diag(U^i,A^i).
\end{equation}
The approach detailed in Section~\ref{sec:magical} yields a unitary
matrix $M$ to realize this linear combination.
On the other hand, we obtain from (\ref{eq:aux}) the relation
$$ f(U)=\sum_{i=0}^{m-1} \alpha_i U^i$$ 
by ignoring the auxiliary matrices $A^i$, $0\le i< m$. 
It is clear that a circuit of
the type shown in Fig.~\ref{fig:final} with $\mu$ chosen such that
$2^{\mu-1}<m\le 2^\mu$ implements this linear combination of the matrices 
$U^i$, $0\le i< m$, provided we
use the matrix $M$ constructed above.~\qed

\mysection{Conclusions} 
\noindent 
Few methods are currently known that facilitate the engineering of
quantum algorithms.  Linear algebra allowed us to derive efficient
quantum circuits for $f(U)$, given an efficient quantum circuit for
$U$, as long as $U^m$ is a scalar matrix for some small integer $m$.
This method can be used in conjuction with the Fourier sampling
techniques by Shor~\cite{Shor:94}, the eigenvalue estimation technique by
Kitaev~\cite{Kitaev:97}, and the probability amplitude amplification
method by Grover~\cite{grover96}, to design more elaborate quantum
algorithms. 

\end{empfile}
\bibliography{paper}
\end{document}